\documentclass[11pt]{article}

\usepackage[margin=1in]{geometry}
\usepackage[utf8]{inputenc}
\usepackage[T1]{fontenc}
\usepackage{microtype}
\usepackage[affil-it]{authblk}
\usepackage{amsmath}
\usepackage{booktabs}
\usepackage{graphicx}
\usepackage[table]{xcolor}
\usepackage{array}
\usepackage{tabularx}
\usepackage{multirow}
\usepackage{caption}
\usepackage{tikz}
\usetikzlibrary{shapes.geometric, arrows.meta, positioning, fit, calc}
\usepackage[round]{natbib}
\usepackage{enumitem}
\usepackage{algorithm}
\usepackage{algpseudocode}
\usepackage{listings}
\usepackage{placeins}

\definecolor{vrBlue}{HTML}{2F5D8C}
\definecolor{vrBlueFill}{HTML}{EAF1F8}
\definecolor{vrTeal}{HTML}{2A7F78}
\definecolor{vrTealFill}{HTML}{E8F4F2}
\definecolor{vrGold}{HTML}{A97724}
\definecolor{vrGoldFill}{HTML}{FBF3E1}
\definecolor{vrWarm}{HTML}{A45345}
\definecolor{vrWarmFill}{HTML}{F7ECE9}
\definecolor{vrInk}{HTML}{252525}
\definecolor{vrGray}{HTML}{66717C}
\definecolor{vrGrayFill}{HTML}{F3F5F7}
\definecolor{vrRule}{HTML}{AAB4BE}
\definecolor{vrHighlight}{HTML}{E4EEF7}

\newcommand{\tableformat}{%
  \renewcommand{\arraystretch}{1.16}%
  \setlength{\tabcolsep}{4.5pt}%
}


\lstdefinestyle{prompt}{
  basicstyle=\footnotesize\ttfamily,
  breaklines=true,
  breakindent=0pt,
  postbreak=\mbox{\textcolor{gray}{$\hookrightarrow$}\space},
  columns=fullflexible,
  keepspaces=true,
  frame=single,
  framesep=5pt,
  rulecolor=\color{black!35},
  xleftmargin=2pt,
  xrightmargin=2pt,
  aboveskip=8pt,
  belowskip=6pt,
}
\usepackage[hidelinks]{hyperref}

\title{\textbf{VecTree-RAG: An Agentic Retrieval-Augmented Generation Framework Combining Vector and Tree Retrieval for Efficiency and Accuracy}}

\author[1]{Xinyan Zhong}
\author[1]{Yuwei Shi}
\author[1]{Yuqi Wei}
\author[4]{Chen Shen}
\author[3]{Tianhang Zhou}
\author[1,2]{Zhenghao Wu\thanks{Email: zhenghao.wu@xjtlu.edu.cn}}
\affil[1]{Department of Chemistry and Materials Science, Xi’an Jiaotong-Liverpool University, Suzhou 215123, Jiangsu, P. R. China}
\affil[2]{Department of Chemistry, University of Liverpool, Liverpool L69 7ZD, United Kingdom}
\affil[3]{College of Carbon Neutrality Future Technology, State Key Laboratory of Heavy Oil Processing, China University of Petroleum (Beijing), Beijing 102249, China}
\affil[4]{Suzhou Lab, Suzhou, P. R. China}

\date{}

\begin{document}

\maketitle

\begin{abstract}
Scientific question answering requires a retrieval system to solve two distinct problems: identifying which papers are relevant and locating the supporting evidence within those papers. Conventional retrieval-augmented generation typically addresses both through similarity search over fixed-length passages, flattening document structure and separating scientific claims from their methodological and argumentative context. We present VecTree-RAG, an agentic framework that assigns these tasks to complementary retrieval mechanisms. Vector search ranks compact document and section representations across the corpus, whereas reasoning-guided traversal of source-verified section trees localizes evidence within shortlisted papers. Full text is retained in a page store and exposed progressively only after structural localization. We evaluate VecTree-RAG on 300 QASPER questions, an open-access subset of 54 LitQA2 questions, and 49 multi-document MOSAIC questions. Compared with Dense RAG, reranked Dense RAG, RAPTOR, and Search-o1, VecTree-RAG obtained the highest observed answer score on all three benchmarks, reaching 0.800 LLM-judge correctness on QASPER, 0.925 accuracy on LitQA2, and a 0.547 composite score on MOSAIC. On QASPER, its evidence-page precision was 0.274, compared with 0.046--0.071 for the baselines. LitQA2 ablations further showed that the complete vector--tree architecture required fewer inference tokens than variants without tree navigation or corpus-level vector routing. These results indicate that vector retrieval narrows the corpus-level search space and tree navigation concentrates reading on structurally relevant evidence. Although multi-turn inference remains more expensive than single-call retrieval, VecTree-RAG provides a structure-aware and traceable architecture for scientific literature question answering.
\end{abstract}


\section{Introduction}
\label{sec:intro}

Retrieval-augmented generation (RAG) commonly represents a document collection as fixed-length passages stored in a vector index \citep{lewis2020rag,gao2023ragsurvey}. This representation is convenient, but it discards the section hierarchy that organizes a scientific paper. Consequently, retrieved passages can be separated from the methods, figures, or qualifications needed to interpret them. The number of indexed vectors also grows with the number of passages, even when most queries require only a small region of each paper.

Retrieving more passages does not necessarily resolve this problem. Expanding the retrieved set increases redundancy and can obscure useful evidence \citep{liu2024lost}, while flat passages provide no representation of how results, methods, and qualifications relate within a paper. These limitations motivate a retrieval architecture that exposes document structure before full text.

VecTree-RAG is an agentic RAG framework that combines two complementary retrieval mechanisms. \emph{Vector retrieval} ranks compact document and section representations across the corpus using semantic and lexical signals, whereas \emph{tree retrieval} navigates the recovered section hierarchy within shortlisted documents to localize pages for closer reading. This \emph{locate-then-read} design treats a paper's authored structure as a retrieval interface rather than flattening it into independent passages. Here, efficiency denotes a compact structural index and selective access to source text; cumulative multi-turn inference cost is evaluated separately in Appendix~\ref{app:inference-cost}.

This paper makes three contributions.
\begin{itemize}[leftmargin=1.4em, itemsep=2pt, topsep=2pt]
  \item We introduce VecTree-RAG, an agentic RAG framework built on the insight that corpus-level paper discovery and within-paper evidence localization require different retrieval mechanisms. The framework assigns semantic ranking across documents to vector retrieval and structure-aware navigation within documents to tree retrieval, coordinated by an agent through a locate-then-read workflow (Section~\ref{sec:method}).
  \item We develop a source-verified hierarchical index and progressive-disclosure retrieval interface that links document and section representations to their original pages, enabling compact indexing and traceable evidence localization without a passage-level vector store. The interface retains page-resolved source evidence to support sentence-level citations in generated answers (Sections~\ref{sec:structure} and~\ref{sec:interface}).
  \item We introduce MOSAIC (Multi-dOcument Scientific Aggregation, Inference, and Comparison), a benchmark dataset for multi-paper scientific synthesis with 49 free-text questions over 124 papers. Each question requires evidence from five to seven papers and tests aggregation, comparison, or multi-hop inference, with reference answers and document--page evidence for evaluating answer quality and evidence localization (Section~\ref{sec:benchmarks}).
\end{itemize}

\section{Related Work}
\label{sec:related}

\paragraph{Chunk-based retrieval-augmented generation.}
The standard RAG pipeline couples a dense retriever \citep{karpukhin2020dpr,khattab2020colbert} with a generator over fixed-size passages \citep{lewis2020rag}. Hybrid retrieval combines dense and lexical signals such as BM25 \citep{robertson2009bm25}, often through reciprocal rank fusion \citep{cormack2009rrf}. Query transformations such as HyDE can further improve zero-shot retrieval \citep{gao2023hyde}. VecTree-RAG applies dense--sparse fusion to document and section representations, but does not embed raw passages. Its index footprint therefore depends on the number of structural entries rather than the number of text chunks.

\paragraph{Hierarchical retrieval and document navigation.}
RAPTOR \citep{sarthi2024raptor} constructs an induced hierarchy by recursively clustering embedded text segments and summarizing the text in each cluster, then retrieves nodes by similarity. MemWalker \citep{chen2023memwalker} treats long-document reading as traversal over a hierarchically summarized memory tree. PageIndex is closely related in retrieval philosophy: its initial design navigates pages and sections through LLM reasoning without a passage-level embedding store \citep{zhang2025pageindex}, and its file-system extension applies hierarchical reasoning to corpus-level routing \citep{pageindex2026filesystem}. VecTree-RAG differs in how it divides retrieval across levels. Vector retrieval performs cross-document paper discovery, whereas tree retrieval navigates the recovered source-authored section hierarchy within shortlisted papers. The hierarchy is grounded in source pages and exposed through a progressive-disclosure interface that retains page-resolved evidence.

\paragraph{Agentic and iterative retrieval.}
Interleaving retrieval with reasoning can improve multi-hop question answering \citep{trivedi2023ircot,press2023selfask}. ReAct-style tool use \citep{yao2023react} generalizes this strategy to broader action spaces. Self-RAG \citep{asai2024selfrag} learns when to retrieve and critique generated responses. PaperQA \citep{lala2023paperqa} gathers scientific evidence through chunk-level search, whereas Search-o1 \citep{li2025searcho1} performs iterative search over flat passages. VecTree-RAG instead exposes search results, section summaries, and page content as progressively larger payloads.

\paragraph{Scientific document QA.}
In scientific-document question answering, QASPER \citep{dasigi2021qasper} established information-seeking QA over full research papers and identified evidence selection as a central bottleneck. SPECTER \citep{cohan2020specter} and SPECTER2 \citep{singh2023scirepeval} provide document representations trained using citation structure. VecTree-RAG uses these representations for cross-document retrieval over titles, descriptions, and section summaries rather than raw passages.

\section{Method}
\label{sec:method}

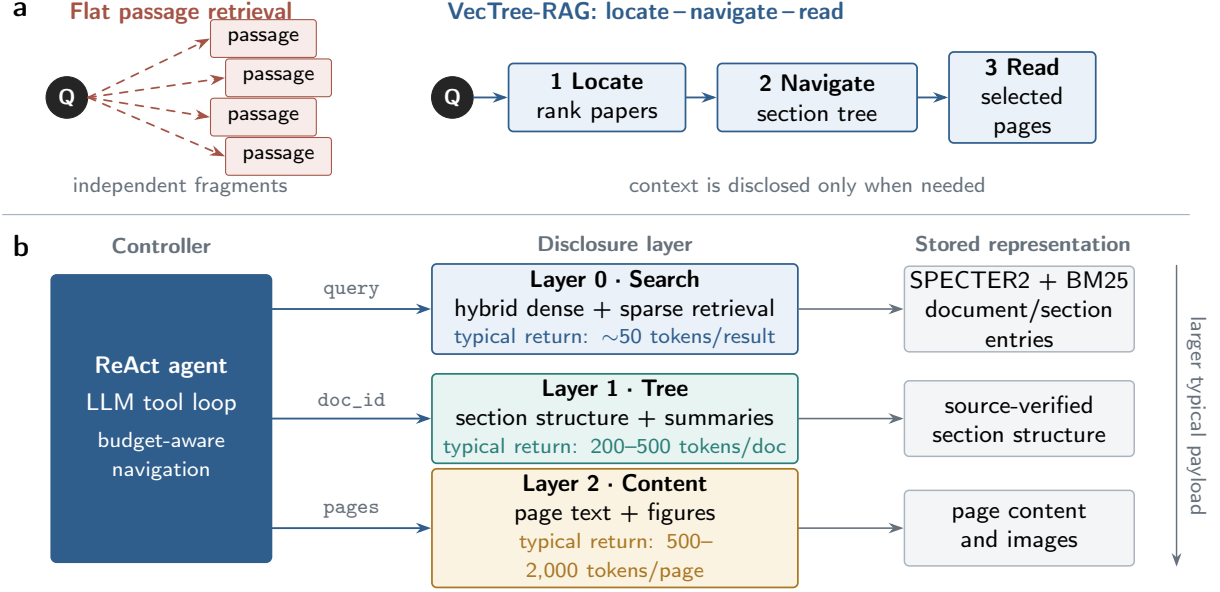
\begin{figure}[t]
\centering
\begin{tikzpicture}[
  x=1cm,
  y=1cm,
  font=\sffamily\footnotesize,
  line cap=round,
  line join=round,
  question/.style={
    draw=vrInk,
    circle,
    minimum size=5.5mm,
    inner sep=0pt,
    fill=vrInk,
    text=white,
    font=\sffamily\bfseries\scriptsize
  },
  chunk/.style={
    draw=vrWarm,
    fill=vrWarmFill,
    rounded corners=1.2pt,
    minimum width=14mm,
    minimum height=5mm,
    inner sep=1.5pt,
    font=\sffamily\scriptsize
  },
  stage/.style={
    draw=vrBlue,
    fill=vrBlueFill,
    line width=.65pt,
    rounded corners=1.8pt,
    align=center,
    minimum height=9mm,
    inner sep=2pt
  },
  layerzero/.style={
    draw=vrBlue,
    fill=vrBlueFill,
    line width=.65pt,
    rounded corners=1.8pt,
    align=center,
    text width=47mm,
    minimum height=10mm,
    inner sep=2pt
  },
  layerone/.style={
    draw=vrTeal,
    fill=vrTealFill,
    line width=.65pt,
    rounded corners=1.8pt,
    align=center,
    text width=47mm,
    minimum height=10mm,
    inner sep=2pt
  },
  layertwo/.style={
    draw=vrGold,
    fill=vrGoldFill,
    line width=.65pt,
    rounded corners=1.8pt,
    align=center,
    text width=47mm,
    minimum height=10mm,
    inner sep=2pt
  },
  store/.style={
    draw=vrRule,
    fill=vrGrayFill,
    line width=.55pt,
    rounded corners=1.8pt,
    align=center,
    text width=29mm,
    minimum height=10mm,
    inner sep=2pt
  },
  agent/.style={
    draw=vrBlue,
    fill=vrBlue,
    text=white,
    line width=.7pt,
    rounded corners=2pt,
    align=center,
    text width=27mm,
    minimum height=38mm,
    inner sep=3pt
  },
  coolarrow/.style={
    -{Stealth[length=2mm,width=1.4mm]},
    draw=vrBlue,
    line width=.7pt
  },
  warmarrow/.style={
    -{Stealth[length=2mm,width=1.4mm]},
    draw=vrWarm,
    line width=.65pt,
    dashed
  },
  neutralarrow/.style={
    -{Stealth[length=2mm,width=1.4mm]},
    draw=vrGray,
    line width=.6pt
  }
]
\path[use as bounding box] (0,-6.35) rectangle (15.7,1.45);

\node[anchor=west, font=\sffamily\bfseries] at (0,1.16) {a};
\node[anchor=west, font=\sffamily\bfseries\footnotesize, text=vrWarm] at (0.75,1.12)
  {Flat passage retrieval};
\node[anchor=west, font=\sffamily\bfseries\footnotesize, text=vrBlue] at (5.75,1.12)
  {VecTree-RAG: locate\,--\,navigate\,--\,read};

\node[question] (flatq) at (0.85,0) {Q};
\node[chunk] (c1) at (3.45,0.78) {passage};
\node[chunk] (c2) at (3.65,0.26) {passage};
\node[chunk] (c3) at (3.45,-0.26) {passage};
\node[chunk] (c4) at (3.65,-0.78) {passage};
\foreach \c in {c1,c2,c3,c4}
  \draw[warmarrow] (flatq.east) -- (\c.west);
\node[font=\sffamily\scriptsize, text=vrGray, align=center] at (2.35,-1.18)
  {independent fragments};

\node[question] (treeq) at (5.95,0) {Q};
\node[stage, text width=22mm, right=4.5mm of treeq] (papers)
  {\textbf{1 Locate}\\rank papers};
\node[stage, text width=25mm, right=4mm of papers] (sections)
  {\textbf{2 Navigate}\\section tree};
\node[stage, text width=18mm, right=4mm of sections] (content)
  {\textbf{3 Read}\\selected pages};
\draw[coolarrow] (treeq.east) -- (papers.west);
\draw[coolarrow] (papers.east) -- (sections.west);
\draw[coolarrow] (sections.east) -- (content.west);
\node[font=\sffamily\scriptsize, text=vrGray, align=center] at (10.65,-1.18)
  {context is disclosed only when needed};

\draw[vrRule, line width=.45pt] (0,-1.55) -- (15.7,-1.55);

\node[anchor=west, font=\sffamily\bfseries] at (0,-1.96) {b};
\node[font=\sffamily\bfseries\scriptsize, text=vrGray] at (2.1,-1.96) {Controller};
\node[font=\sffamily\bfseries\scriptsize, text=vrGray] at (8.1,-1.96) {Disclosure layer};
\node[font=\sffamily\bfseries\scriptsize, text=vrGray] at (13.5,-1.96) {Stored representation};

\node[agent] (agentbox) at (2.1,-4.25)
  {\textbf{ReAct agent}\\[1mm]LLM tool loop\\[1mm]\scriptsize budget-aware navigation};

\node[layerzero] (l0) at (8.1,-2.8)
  {\textbf{Layer 0 · Search}\\hybrid dense + sparse retrieval\\[-.2mm]
   {\scriptsize\color{vrBlue} typical return: $\sim$50 tokens/result}};
\node[layerone] (l1) at (8.1,-4.25)
  {\textbf{Layer 1 · Tree}\\section structure + summaries\\[-.2mm]
   {\scriptsize\color{vrTeal} typical return: 200--500 tokens/doc}};
\node[layertwo] (l2) at (8.1,-5.70)
  {\textbf{Layer 2 · Content}\\page text + figures\\[-.2mm]
   {\scriptsize\color{vrGold} typical return: 500--2{,}000 tokens/page}};

\node[store] (s0) at (13.45,-2.8) {SPECTER2 + BM25\\document/section entries};
\node[store] (s1) at (13.45,-4.25) {source-verified\\section structure};
\node[store] (s2) at (13.45,-5.70) {page content\\and images};

\draw[coolarrow] (agentbox.east |- l0.west) -- (l0.west)
  node[midway, above, font=\sffamily\ttfamily\scriptsize, text=vrGray] {query};
\draw[coolarrow] (agentbox.east |- l1.west) -- (l1.west)
  node[midway, above, font=\sffamily\ttfamily\scriptsize, text=vrGray] {doc\_id};
\draw[coolarrow] (agentbox.east |- l2.west) -- (l2.west)
  node[midway, above, font=\sffamily\ttfamily\scriptsize, text=vrGray] {pages};

\draw[neutralarrow] (l0.east) -- (s0.west);
\draw[neutralarrow] (l1.east) -- (s1.west);
\draw[neutralarrow] (l2.east) -- (s2.west);

\coordinate (payloadtop) at ([xshift=5.5mm]s0.north east);
\coordinate (payloadbottom) at (payloadtop |- s2.south);
\draw[-{Stealth[length=1.8mm,width=1.3mm]}, draw=vrGray, line width=.55pt]
  (payloadtop) -- (payloadbottom)
  node[midway, rotate=-90, above, font=\sffamily\scriptsize, text=vrGray]
  {larger typical payload};
\end{tikzpicture}
\caption{\textbf{Architecture and progressive disclosure in VecTree-RAG.}
\textbf{a,} Flat RAG returns independent passages. VecTree-RAG instead ranks papers, navigates a source-verified section structure, and reads selected pages.
\textbf{b,} The corpus agent moves from compact search results to section summaries and then page text or figures; the right column shows the corresponding stores. Token ranges denote typical serialized tool returns (design estimates), not total model context or fixed costs. Only document and section representations are embedded.}
\label{fig:architecture}
\end{figure}

VecTree-RAG follows a locate-then-read design (Fig.~\ref{fig:architecture}). The index preserves the section structure recovered from each paper. Vector similarity ranks compact cross-document representations, while within-document localization uses titles and summaries. The progressive-disclosure interface exposes search results, section structures, and page content in increasing payload sizes.

VecTree-RAG separates document ingestion from query-time retrieval. During ingestion, each PDF is converted into a source-verified section structure; compact document and section representations are indexed, while page text and figures are retained in separate source stores (Section~\ref{sec:structure}). At query time, the evaluated agent uses seven tools across three progressive-disclosure layers (Sections~\ref{sec:interface}--\ref{sec:tools}).

\subsection{Preliminaries: Agentic Search}
\label{sec:prelim}

The query-time controller is a ReAct-style agent \citep{yao2023react} that interleaves reasoning and tool use. Given a question $q$, the interaction runs for at most $L$ rounds. At round $\ell$ the state $s_\ell$ is the message history: the system prompt, the query, and the trajectory of prior tool calls and observations. The policy samples an action
\begin{equation}
a_\ell \sim \pi_\theta(\cdot \mid s_\ell), \qquad a_\ell \in \{\textsc{Final}\} \cup \mathcal{A},
\end{equation}
where an action is either a final answer or a tool invocation $a_\ell = (\textit{name}, \mathbf{x})$ with arguments $\mathbf{x}$. Executing a tool yields an observation $o_\ell$, and the state grows by concatenation,
\begin{equation}
s_{\ell+1} \leftarrow s_\ell \oplus (a_\ell, o_\ell).
\end{equation}
In VecTree-RAG the action set $\mathcal{A}$ comprises seven tools spanning three disclosure layers. Layer~0 provides \texttt{search\_documents} and \texttt{get\_paper\_metadata}; Layer~1 provides \texttt{get\_document\_tree} and \texttt{search\_within\_paper}; and Layer~2 provides \texttt{get\_page\_content}, \texttt{list\_images}, and \texttt{get\_page\_image}. Sections~\ref{sec:interface}--\ref{sec:tools} describe their roles in the locate-then-read workflow.

\subsection{Document Structure Modeling and Verified Indexing}
\label{sec:structure}

We model each document $d$ as its native section hierarchy, a tree $T_d$ whose nodes are the document's own sections. A node is
\begin{equation}
\nu = \big(\textit{id}(\nu),\ \textit{title}(\nu),\ \textit{summary}(\nu),\ [\,p^{\uparrow}(\nu), p^{\downarrow}(\nu)\,]\big),
\end{equation}
where $[\,p^{\uparrow}, p^{\downarrow}\,]$ is the inclusive page range that grounds the node in the parsed source. Indexing produces three artifacts: the tree $T_d$, a page-contents file, and a directory of extracted figures. The tree contains hierarchy, summaries, and metadata, but no body text. The LLM-assisted stages are paired with deterministic checks where source-grounded validation is available.

\paragraph{Parsing.}
PDFs are converted to per-page markdown with the Marker parser \citep{paruchuri2024marker}, which preserves page boundaries and extracts embedded images with page numbers and bounding boxes. Page indices are the coordinate system for everything downstream: every tree node ultimately resolves to an inclusive page range.

\paragraph{Structure extraction with verification.}
An LLM reads overlapping page groups, each bounded at 20{,}000 tokens with one page of overlap. It emits entries $\{(s_i,t_i,p_i)\}$, where $s_i$ is a dotted index, $t_i$ is a title, and $p_i$ is a start page. Each title must match the claimed page within a $\pm1$-page tolerance using a case-insensitive substring rule. We call the fraction of entries passing this rule the \emph{section-title verification rate}. This measure checks title occurrence and approximate location; it does not validate summary faithfulness or the complete hierarchy. Failed entries are re-extracted from focused candidate pages for at most three repair rounds. A second LLM pass determines whether a title begins its start page, which resolves pages shared by adjacent sections. Finally, a deterministic scan inserts an abstract node when needed.

\paragraph{Tree construction and bounded splitting.}
The verified flat list is folded into a tree using the dotted indices. Each page range is computed from the node's start page and the next verified heading in document order, before the hierarchy is constructed. Nodes exceeding \emph{both} 20{,}000 tokens and 10 pages are split by recursively rerunning structure extraction on the node text. This rule bounds nodes that exceed both thresholds, but it does not guarantee a universal maximum on either dimension. Nodes receive stable zero-padded identifiers in depth-first order.

\paragraph{Enrichment and metadata.}
Each node is summarized from its section text. Parent summaries use preamble text, whereas leaf summaries use the complete body. Sections shorter than 200 tokens retain their source text instead of an LLM summary.

At document level, the pipeline produces a one-sentence description, domain keywords, and bibliographic metadata. Metadata processing begins with regular-expression extraction of DOI, year, title, journal, volume, and pages. An LLM then extracts fields using these candidates as hints. Finally, Crossref verifies and enriches available fields under explicit precedence and format-validation rules. Enrichment calls run concurrently under a semaphore.

The resulting tree contains titles, page ranges, identifiers, and summaries, but no body text. In the current interface, a serialized tree is estimated to return approximately 200--500 tokens for a typical paper. We treat this range as a design estimate because the evaluated corpora do not include a reported distribution of serialized tree sizes.

\subsection{Progressive-Disclosure Interface}
\label{sec:interface}

The query-time controller is a ReAct-style tool loop with an approximately 400-token system prompt. The prompt asks the agent to search before opening documents, inspect tree summaries before requesting pages, and cite page-level evidence. The evaluated loop permits up to 30 iterations. Every model-issued tool call is counted, including repeated, malformed, or failed calls, and a synthesis reminder is appended from the tenth call onward. Page requests are capped at ten pages per call. Image requests return figures as vision inputs when the text layer does not capture their content.

Table~\ref{tab:layers} summarizes the typical payload returned by each layer. These values describe serialized tool outputs rather than total API-billed tokens, which also include prompts, history, and generated text.

\begin{table}[t]
\centering
\small
\tableformat
\begin{tabularx}{\textwidth}{c >{\ttfamily\raggedright\arraybackslash}p{35mm} X >{\raggedleft\arraybackslash}p{34mm}}
\toprule
\rowcolor{vrBlueFill}
\textbf{Layer} & \textbf{\rmfamily Tool} & \textbf{Returns} & \textbf{Payload estimate (tokens)} \\
\midrule
\rowcolor{vrBlueFill!58}
\multirow{2}{*}{\textbf{\color{vrBlue}0}} & search\_documents & metadata + matching sections & $\sim$50 / result \\
\rowcolor{vrBlueFill!58}
 & get\_paper\_metadata & metadata + top sections & $\sim$50 / doc \\
\addlinespace[2pt]
\rowcolor{vrTealFill}
\multirow{2}{*}{\textbf{\color{vrTeal}1}} & get\_document\_tree & section structure + summaries & 200--500 / doc \\
\rowcolor{vrTealFill}
 & search\_within\_paper & matching section titles & query-dependent \\
\addlinespace[2pt]
\rowcolor{vrGoldFill}
\multirow{3}{*}{\textbf{\color{vrGold}2}} & get\_page\_content & full-page markdown ($\leq$10 pages/call) & 500--2{,}000 / page \\
\rowcolor{vrGoldFill}
 & list\_images & available figure filenames & file-count-dependent \\
\rowcolor{vrGoldFill}
 & get\_page\_image & base64 figure (vision input) & model-dependent \\
\bottomrule
\end{tabularx}
\caption{\textbf{Progressive-disclosure tool interface.} Payload estimates refer to serialized tool returns and exclude the accumulated prompt, conversation history, and model output.}
\label{tab:layers}
\end{table}

Algorithm~\ref{alg:loop} states the loop.

\begin{algorithm}[t]
\caption{VecTree-RAG query-time agent loop (locate-then-read).}
\label{alg:loop}
\begin{algorithmic}[1]
\Require Query $q$; corpus index $\mathcal{I}$ (SPECTER2{+}BM25 over doc/section entries); metadata registry $\mathcal{M}$; source-verified structures $\{T_d\}$; page and image stores $\mathcal{P}$; iteration cap $L$
\State $s_1 \gets [\textsc{System},\ \textsc{User}\!:\,q]$
\State $c \gets 0$ \Comment{number of model-issued tool calls}
\For{$\ell = 1$ \textbf{to} $L$}
  \State $a_\ell \sim \pi_\theta(\cdot \mid s_\ell)$
  \If{$a_\ell = \textsc{Final}$} \State \Return answer \EndIf
  \State $c \gets c + 1$
  \If{$a_\ell.\textit{name} = \texttt{search\_documents}$}
    \State $o_\ell \gets \textsc{HybridRetrieve}(a_\ell.\mathbf{x}, \mathcal{I})$ \Comment{Layer~0}
  \ElsIf{$a_\ell.\textit{name} = \texttt{get\_paper\_metadata}$}
    \State $o_\ell \gets \textsc{MetadataLookup}(a_\ell.\textit{doc}, \mathcal{M})$ \Comment{Layer~0}
  \ElsIf{$a_\ell.\textit{name} = \texttt{get\_document\_tree}$}
    \State $o_\ell \gets T_{a_\ell.\textit{doc}}$ \Comment{Layer~1}
  \ElsIf{$a_\ell.\textit{name} = \texttt{search\_within\_paper}$}
    \State $o_\ell \gets \textsc{HybridRetrieve}(a_\ell.\mathbf{x}, \mathcal{I}\mid a_\ell.\textit{doc})$ \Comment{Layer~1}
  \ElsIf{$a_\ell.\textit{name} = \texttt{get\_page\_content}$}
    \State $o_\ell \gets \mathcal{P}[\,a_\ell.\textit{doc},\, a_\ell.\textit{pages}\,]$ \Comment{Layer~2 text}
  \ElsIf{$a_\ell.\textit{name} = \texttt{list\_images}$}
    \State $o_\ell \gets \textsc{ImageList}(a_\ell.\textit{doc}, \mathcal{P})$ \Comment{Layer~2 index}
  \ElsIf{$a_\ell.\textit{name} = \texttt{get\_page\_image}$}
    \State $o_\ell \gets \textsc{ImageRead}(a_\ell.\mathbf{x}, \mathcal{P})$ \Comment{Layer~2 vision}
  \EndIf
  \State \textbf{if} $c \geq 10$ \textbf{then} append a synthesize-now note
  \State $s_{\ell+1} \gets s_\ell \oplus (a_\ell, o_\ell)$
\EndFor
\State \Return most recent nonempty assistant text; otherwise raise an evaluation error
\end{algorithmic}
\end{algorithm}

\subsection{Tools and Synergy}
\label{sec:tools}

\paragraph{Inter-document retrieval (Layer~0).}
The first disclosure layer identifies candidate papers. Each \emph{document entry} contains the title, description, authors, journal, and top-level section titles. Each top-level section contributes a separate \emph{section entry} containing its title and summary. Both entry types are short, abstract-like texts.

Dense scores are cosine similarities between the L2-normalized SPECTER2 embedding (768-d, 512-token input cap) of the query and each entry. Sparse scores come from BM25 over lowercased alphanumeric tokens of the same entries. Per document, the best-scoring entry of each type is taken, so a paper can be surfaced either because it is globally about the query or because one section is. Candidates must pass a pre-fusion relevance gate (dense similarity $\geq 0.35$ or BM25 score $\geq 1.0$) that discards documents which neither retriever considers relevant, rather than letting rank fusion promote them. Surviving candidates are fused by reciprocal rank fusion with the standard constant $k = 60$:
\begin{equation}
\mathrm{RRF}(d) \;=\; \frac{1}{k + r_{\text{dense}}(d)} \;+\; \frac{1}{k + r_{\text{sparse}}(d)},
\end{equation}
where $r_{\cdot}(d)$ is the document's rank under each retriever. A post-fusion gate adapts to corpus size. Above 50 documents, it removes results below 40\% of the leading RRF score. For smaller corpora, it requires retriever agreement or a strong individual match. The strong-match thresholds are 0.55 for dense similarity and 4.0 for BM25. These gates filter marginal results before they are presented to the agent.

Each returned result carries only metadata (identifier, title, authors, venue, year, page count, one-sentence description, and the matching section titles), at roughly 50 tokens. The matching-sections field bridges to the next layer: it tells the agent which paper to open and where in it to look first.

\paragraph{Within-document navigation and standalone verified tree search (Layer~1).}
The evaluated corpus agent loads titles, summaries, and page ranges through \texttt{get\_document\_tree}. It then selects pages directly within its tool loop. This navigation step does not invoke a separate deterministic validation procedure.

The standalone single-document mode uses an additional verified tree-search procedure. An LLM returns candidate node identifiers, which are checked for existence, valid page ranges, and lexical overlap with the query. The fraction of passing nodes defines a rule-based confidence score. Scores below 0.5 trigger at most two retries with an error report. If confidence remains low, a lexical fallback ranks nodes by stopword-filtered overlap and returns the top five. This validation checks output conformity, not whether the selected sections contain the correct answer.

\paragraph{Complexity and index footprint.}
For a corpus of $N$ documents with an average of $S$ indexed section entries per document, the vector index contains $N(1+S)$ embeddings. A chunk store instead contains a number of vectors proportional to total corpus tokens divided by chunk size. We do not report a corpus-matched storage benchmark, so we restrict the claim to this difference in index construction. Query-time cost is dominated by repeated LLM calls and is reported in Appendix~\ref{app:inference-cost}. Indexing requires structure extraction and summarization; Appendix~\ref{app:intrinsics} reports its measured wall time.

\paragraph{Composition.} Layer~0 ranks papers and returns the titles of their best-matching sections. Layer~1 exposes the source-verified section structure for direct agent navigation. Layer~2 provides page text and images for close reading. The ablations in Section~\ref{sec:ablations} test the vector and tree components within the evaluated corpus agent.

\section{Experiments}
\label{sec:experiments}

We evaluate VecTree-RAG in three regimes. QASPER tests question answering within one known paper. LitQA2 tests corpus-level retrieval when one paper contains the answer. MOSAIC tests synthesis across at least five papers. The main evaluation reports answer quality and evidence localization at the granularity available for each benchmark. Cumulative API-billed tokens and wall time are reported in Appendix~\ref{app:inference-cost}.

\subsection{Benchmark Details}
\label{sec:benchmarks}

\paragraph{QASPER (single-document).} We use a stratified sample of 300 questions drawn once with a fixed sampling seed from the QASPER test split \citep{dasigi2021qasper}. The complete ingested split contains 1{,}451 questions across 416 papers. QASPER's supplied section structure is represented with one synthetic page per paragraph. Gold paragraph evidence therefore aligns with the page identifiers used by the agent.

\paragraph{LitQA2 (corpus-level single-source QA).} We use an open-access subset of LitQA2 from LAB-Bench \citep{laurent2024labbench}. Gold DOIs were resolved through OpenAlex, Unpaywall, and Europe PMC, yielding 62 accessible PDFs. Marker parsing and indexing succeeded for 61 papers. After excluding questions without an indexed gold paper, 54 questions remained in a closed 61-paper corpus. This accessibility filter can introduce selection bias relative to the full benchmark.

\paragraph{MOSAIC (multi-document synthesis).} MOSAIC contains 49 free-text questions over 124 materials-science and computational-science papers. Each question requires evidence from five to seven papers. The set includes 24 inference, 13 aggregation, and 12 comparison questions. Each item has a free-text reference answer, an explicit reasoning chain, and gold $(\text{document},\text{page})$ evidence pairs. Every system was evaluated on all 49 questions across three seeds.

\paragraph{MOSAIC construction and quality control.} Candidate paper groups were formed as dynamic topical neighborhoods using SPECTER2 similarity over titles and document descriptions, combined with keyword overlap. DeepSeek-V4-Pro drafted one question, reference answer, reasoning chain, and evidence set at a time from the selected source pages at temperature 0.35. Aggregation questions collect directly stated facts across papers; comparison questions contrast a shared axis; and inference questions require a conclusion not stated by any single paper. Automated checks at temperature 0.0 tested source grounding, answerability, closed-book leakage, multi-paper coverage, and duplication. Inference questions additionally passed single-document sufficiency and chunk-retrieval checks. Failed grounding or coverage checks triggered a bounded repair loop. The released set was constructed and screened automatically without independent human validation or adjudication.

\subsection{Baselines, Systems, and Protocol}
\label{sec:settings}

\paragraph{Systems.} All reported experiments used repository snapshot \texttt{a57c300} and the seven-tool interface described in Section~\ref{sec:tools}. We compare the full VecTree-RAG agent, with a 30-iteration cap, against four baselines used consistently across the three benchmarks. On QASPER and LitQA2, Dense RAG retrieved the top ten 800-token passages with 400-token overlap. On MOSAIC, it used an expanded-context setting with $k=232$ to test whether a large flat context could recover evidence across many papers. This setting is a retrieval-volume stress test rather than an exact match of realized API cost. Dense RAG with reranking rescored 30 dense candidates with a BGE cross-encoder \citep{xiao2024bge} and retained ten. RAPTOR \citep{sarthi2024raptor} retrieved from a recursive summary tree. Search-o1 \citep{li2025searcho1} used a ten-round flat-passage search loop. QASPER retrieval systems received the gold paper, whereas LitQA2 systems searched the complete corpus.

\paragraph{Generation, prompts, and grading.} All answer-generation systems used DeepSeek-V4-Flash through OpenRouter at temperature 1.0. Each system was evaluated with three generation seeds, and reported metrics are means over the three seed-level results. DeepSeek-V4-Pro at temperature 0.0 judged every system under a shared rubric. On QASPER, it assessed correctness against the official answers; this metric is distinct from official token-level F1. On LitQA2, it assessed whether the response selected the correct option. Available percentile-bootstrap intervals quantify variation across questions.

\subsection{Evaluation Metrics}
\label{sec:metrics}

All metrics are computed per question and macro-averaged within each seed; tables report the mean of the three seed-level values. QASPER correctness is the proportion of responses assigned \emph{correct} by the judge. A \emph{partial} judgment receives zero in this strict headline metric. LitQA2 accuracy is the proportion of responses that select the gold option.

Evidence precision and recall are computed from sets, so repeated reads of the same evidence unit are counted once. QASPER uses paragraph-level synthetic page identifiers and scores against the best-matching annotator evidence set. A retrieved baseline chunk inherits every paragraph identifier that it spans. MOSAIC instead uses exact $(\text{document},\text{page})$ pairs. LitQA2 key passages are mapped to physical pages by source-text matching; only 10 of 54 questions have a nonempty mapped gold-page set. We therefore treat LitQA2 page scores as an exploratory mapping diagnostic rather than a corpus-level retrieval metric.

For MOSAIC, the judge decomposes an answer into atomic statements and compares them with the reference statements to obtain factuality precision, recall, and F1. Coverage is the fraction of reference facts attributed to the response, and faithfulness is the fraction of response statements judged inferable from the retrieved context. Semantic similarity uses \texttt{allenai/specter2\_base} with a 512-token input cap, the 768-dimensional \texttt{[CLS]} representation, and L2 normalization. Cosine similarity is scaled to $s_{\mathrm{cos}}=(\cos+1)/2$, and the reported composite is
\begin{equation}
\mathrm{Composite}=0.75\,\mathrm{F1}_{\mathrm{factuality}}+0.25\,s_{\mathrm{cos}}.
\end{equation}

\subsection{Main Results}
\label{sec:results}

\paragraph{Single-document QASPER.} VecTree-RAG obtained a mean LLM-judge correctness rate of 0.800 across three seeds on the 300-question sample, higher than all four evaluated baselines (Table~\ref{tab:main-accuracy}). The corresponding scores were 0.750 for Dense RAG, 0.757 for Dense RAG with reranking, 0.690 for RAPTOR, and 0.757 for Search-o1. VecTree-RAG also reached evidence-page precision of 0.274, compared with 0.046--0.071 for these baselines, while retaining evidence recall of 0.737 (Table~\ref{tab:evidence-localization}). Thus, its accuracy advantage was accompanied by a more concentrated set of source evidence.

\begin{table}[t]
\centering
\small
\tableformat
\begin{tabular}{lccc}
\toprule
\rowcolor{vrBlueFill}
\textbf{System} & \textbf{QASPER correctness} $\uparrow$ & \textbf{LitQA2 accuracy} $\uparrow$ & \textbf{MOSAIC composite} $\uparrow$ \\
\midrule
Dense RAG & 0.750 & 0.870 & 0.503 \\
Dense RAG w/ Reranker & 0.757 & 0.889 & 0.334 \\
RAPTOR & 0.690 & 0.815 & 0.329 \\
Search-o1 & 0.757 & 0.759 & 0.489 \\
\addlinespace[2pt]
\rowcolor{vrHighlight}
VecTree-RAG (agent) & \textbf{0.800} & \textbf{0.925} & \textbf{0.547} \\
\bottomrule
\end{tabular}
\caption{\textbf{Benchmark-specific answer scores.} Values are means over three seeds on fixed sets of 300 QASPER, 54 LitQA2, and 49 MOSAIC questions. QASPER and LitQA2 use DeepSeek-V4-Pro judgments at temperature 0.0, whereas MOSAIC uses $0.75$ factuality-F1 $+\,0.25$ scaled SPECTER cosine. The MOSAIC Dense RAG entry uses the expanded-context setting with $k=232$; QASPER and LitQA2 use $k=10$. The metrics are not commensurate and are therefore not averaged across benchmarks.}
\label{tab:main-accuracy}
\end{table}

\begin{table}[t]
\centering
\small
\tableformat
\begin{tabular}{lcc}
\toprule
\rowcolor{vrBlueFill}
\textbf{System} & \textbf{Evidence recall} $\uparrow$ & \textbf{Evidence precision} $\uparrow$ \\
\midrule
Dense RAG & 0.884 & 0.048 \\
Dense RAG w/ Reranker & 0.894 & 0.048 \\
RAPTOR & \textbf{0.899} & 0.046 \\
Search-o1 & 0.817 & 0.071 \\
\addlinespace[2pt]
\rowcolor{vrHighlight}
VecTree-RAG (agent) & 0.737 & \textbf{0.274} \\
\bottomrule
\end{tabular}
\caption{\textbf{QASPER evidence localization.} Values are means over three seeds. Recall and precision compare deduplicated evidence sets with gold paragraph identifiers. Retrieved baseline chunks inherit all synthetic identifiers that they span. The values therefore measure paragraph-level evidence selection rather than localization to physical PDF pages.}
\label{tab:evidence-localization}
\end{table}

\paragraph{Corpus-level LitQA2.} VecTree-RAG achieved a mean accuracy of 0.925 across three seeds (95\% question-bootstrap CI, 0.85--0.98), higher than all four evaluated baselines. The corresponding scores were 0.870 for Dense RAG, 0.889 for Dense RAG with reranking, 0.815 for RAPTOR, and 0.759 for Search-o1 (Table~\ref{tab:main-accuracy}). The result shows that the vector--tree agent can preserve high answer accuracy while locating the relevant source within a 61-paper corpus. Query-time computation is reported in Appendix~\ref{app:inference-cost}.

\paragraph{MOSAIC.} The MOSAIC composite combines factuality-F1 with scaled SPECTER similarity; coverage and faithfulness use the same judge family. Evidence metrics require the correct document and page. VecTree-RAG achieved a composite score of 0.547 and coverage of 0.532, both higher than the same four evaluated baselines (Table~\ref{tab:mosaic}). It also obtained the highest observed evidence precision, 0.221, while maintaining evidence recall of 0.586. Expanded-context Dense RAG reached higher evidence recall, 0.661, but with precision of 0.019. Thus, in the multi-document setting, VecTree-RAG combined strong synthesis quality with a substantially more concentrated set of document--page evidence.

\begin{table}[t]
\centering
\small
\tableformat
\setlength{\tabcolsep}{3.8pt}
\begin{tabular}{lrrrrr}
\toprule
\rowcolor{vrBlueFill}
\textbf{System} & \textbf{Composite} $\uparrow$ & \textbf{Coverage} $\uparrow$ & \textbf{Faith} $\uparrow$ & \textbf{Ev.R} $\uparrow$ & \textbf{Ev.P} $\uparrow$ \\
\midrule
Dense RAG (expanded, $k=232$) & 0.503 & 0.294 & 0.033 & \textbf{0.661} & 0.019 \\
Dense RAG w/ Reranker & 0.334 & 0.114 & \textbf{0.174} & 0.153 & 0.076 \\
RAPTOR & 0.329 & 0.073 & 0.097 & 0.404 & 0.050 \\
Search-o1 & 0.489 & 0.318 & 0.134 & 0.411 & 0.086 \\
\addlinespace[2pt]
\rowcolor{vrHighlight}
VecTree-RAG (agent) & \textbf{0.547} & \textbf{0.532} & 0.170 & 0.586 & \textbf{0.221} \\
\bottomrule
\end{tabular}
\caption{\textbf{MOSAIC multi-document synthesis.} Values are means over three seeds on all 49 questions. The composite is $0.75$ factuality-F1 $+\,0.25$ scaled SPECTER cosine. Ev.R and Ev.P denote document-and-page-aware evidence recall and precision. Dense RAG uses an expanded-context setting with top-$k=232$; this is not an exact match of realized API cost.}
\label{tab:mosaic}
\end{table}

\subsection{Source-Traceable Evidence Localization}
\label{sec:behavior}

VecTree-RAG read a smaller fraction of irrelevant QASPER pages, as reflected by evidence precision of 0.274 compared with 0.046--0.071 for the four retrieval baselines. The corresponding recall of 0.737 was lower than the baseline range of 0.817--0.899. The method therefore improved the concentration of relevant evidence rather than complete evidence coverage. Because each selected item resolves to a source page or QASPER paragraph identifier, the resulting evidence set can be inspected directly against the original document.

Verifiability operates at two bounded levels. During indexing, deterministic checks confirm section-title occurrence and approximate source location. During evaluation, evidence metrics compare deduplicated retrieved units with benchmark-native gold evidence. These checks make the retrieval basis traceable, but they do not establish the faithfulness of generated summaries or the factual correctness of every answer. The harness preserves the set of read pages but not the ordered tool trajectory, so adherence to the intended layer order cannot yet be quantified.

\FloatBarrier
\subsection{Component Ablations}
\label{sec:ablations}

On LitQA2, we remove the tree layer, vector layer, or relevance gates from the corpus agent. The standalone verify--retry--fallback procedure is characterized separately in Appendix~\ref{app:intrinsics}. Progressive disclosure itself was not isolated as an ablation.

\begin{table}[ht]
\centering
\small
\tableformat
\begin{tabular}{lc}
\toprule
\rowcolor{vrBlueFill}
\textbf{Variant (agent)} & \textbf{MCQ accuracy} $\uparrow$ {\scriptsize[95\% CI]} \\
\midrule
\rowcolor{vrHighlight}
Full system & \textbf{0.925}~{\scriptsize[0.85, 0.98]} \\
\addlinespace[2pt]
\rowcolor{vrGrayFill}
$-$ tree layer (search $\to$ pages) & 0.904~{\scriptsize[0.83, 0.98]} \\
$-$ vector layer (tree only, all docs) & 0.811~{\scriptsize[0.70, 0.91]} \\
\rowcolor{vrGrayFill}
$-$ relevance gates (raw RRF) & 0.906~{\scriptsize[0.81, 0.98]} \\
\bottomrule
\end{tabular}
\caption{\textbf{LitQA2 component ablations.} Accuracy values are means over three seeds; brackets show question-bootstrap 95\% CIs. The table reports descriptive component comparisons without inferential $p$-values. Limited page-mapping diagnostics are reported in Appendix~\ref{app:litqa-evidence}.}
\label{tab:ablations}
\end{table}

\FloatBarrier
\section{Discussion and Limitations}
\label{sec:discussion}

Across three scientific-document settings, VecTree-RAG achieved the highest observed answer score relative to Dense RAG, Dense RAG with reranking, RAPTOR, and Search-o1. It reached 0.800 LLM-judge correctness on QASPER, 0.925 accuracy on LitQA2, and a 0.547 MOSAIC composite score. The advantage was accompanied by more selective evidence localization: QASPER evidence precision was 0.274, compared with 0.046--0.071 for the four baselines, while MOSAIC evidence precision reached 0.221. These results indicate that combining cross-document vector retrieval with within-document tree retrieval can improve observed answer quality while concentrating retrieval on source-relevant pages.

Reasoning-based hierarchical systems such as MemWalker and PageIndex are attractive for within-document navigation because an LLM can inspect textual section representations and decide which branch to read \citep{chen2023memwalker,zhang2025pageindex}. PageIndex has also described a query-dependent file-level hierarchy that extends tree search to corpus-level routing \citep{pageindex2026filesystem}. This development highlights an architectural choice: LLM traversal can exploit an informative hierarchy, but broad or weakly labeled branches may require larger routing views or additional reasoning steps. VecTree-RAG takes a complementary design point, using vector retrieval to rank compact document and section representations across the corpus and reserving tree reasoning for shortlisted papers. This specialization limits LLM-based structural navigation to a small candidate set while retaining page-resolved source text for sentence-level citations.

The LitQA2 ablation provides internal evidence for this architectural specialization. When vector routing was removed and the agent navigated trees across all 61 papers, mean cumulative API-billed tokens increased from 127{,}872 to 490{,}482 per completed query, while observed accuracy decreased from 0.925 to 0.811. Removing tree navigation also increased token use to 194{,}821. These results suggest that vector routing is particularly important for reducing the corpus-level reasoning space, whereas tree navigation provides a more economical interface to source text after document selection. Removing the relevance gates reduced token use to 104{,}215 and produced a similar observed accuracy of 0.906, so the gates should be viewed as routing heuristics rather than an independently established efficiency gain. At the index level, VecTree-RAG stores embeddings for document and section representations rather than fixed-length passages. Its vector index therefore scales with structural entries rather than corpus chunk count.

The efficiency claim is therefore scoped to index representation and navigation within the agentic design; it does not imply uniformly lower end-to-end inference cost. The multi-turn agent used more query-time tokens than Dense RAG, Dense RAG with reranking, and RAPTOR; relative to Search-o1, its cost was higher on QASPER but lower on LitQA2 and MOSAIC. Detailed measurements are reported in Appendix~\ref{app:inference-cost}. A fuller efficiency comparison should jointly account for embedding counts, serialized index size, indexing cost, retrieval latency, and cached-prefix tokens. Smaller navigation models, prefix caching, and context compaction may reduce query-time computation while preserving evidence traceability.

VecTree-RAG also has method-level boundaries. It assumes informative section structure and may transfer poorly to unstructured documents. Navigation depends on generated summaries whose faithfulness is not currently verified, and page-level grounding succeeds only when the agent chooses the relevant section. Page-resolved sources can support sentence-level citations, but the present evaluation measures evidence localization only at paragraph or page granularity. MOSAIC was generated and screened automatically without independent human validation, and its construction and scoring use the same model family. Future work should therefore add expert adjudication, validate generated section summaries, record ordered tool trajectories, and test the method on larger and less regularly structured corpora.

\section{Conclusion}
\label{sec:conclusion}

VecTree-RAG assigns corpus-level paper discovery to vector retrieval and within-paper evidence localization to tree retrieval. Against Dense RAG, Dense RAG with reranking, RAPTOR, and Search-o1, it achieved the highest observed answer score on QASPER (0.800), LitQA2 (0.925), and MOSAIC (0.547). It also produced substantially higher QASPER evidence precision (0.274 versus 0.046--0.071) and the highest observed MOSAIC evidence precision (0.221). The source-verified hierarchy links compact document and section representations to original pages, providing an inspectable basis for page-resolved citations. These findings support complementary retrieval mechanisms for accurate and traceable scientific question answering.

\section*{Code Availability}
The source code, evaluation harness, and dataset-construction pipeline are available in the \href{https://github.com/Chenghao-Wu/VecTree-RAG}{VecTree-RAG repository}.

\FloatBarrier
\appendix
\section{Dataset and Corpus Statistics}
\label{app:stats}
Table~\ref{tab:stats} reports the question and corpus statistics for the three evaluation regimes. The QASPER study uses a stratified sample of 300 questions drawn once with a fixed sampling seed from the 1{,}451-question test split (416 papers); the sample touches 215 distinct gold papers and preserves the official answer-type proportions (extractive 50.3\%, abstractive 25.7\%, yes/no 14.3\%, unanswerable 9.7\%). Because QASPER papers are ingested with one synthetic page per paragraph (Section~\ref{sec:benchmarks}), the ``pages'' column for QASPER counts paragraphs rather than physical pages and is therefore not comparable to the two PDF-parsed corpora. LitQA2 and MOSAIC page counts are physical pages recovered by Marker and read directly from each tree's \texttt{total\_pages} field. The evidence row records benchmark-native annotations: gold paragraphs for QASPER, one supplied key passage for each LitQA2 question, and document-level evidence records with page locations for MOSAIC. Only 10 LitQA2 key passages were matched to physical pages, so its mapped-page statistics are reported separately as an exploratory diagnostic.

\begin{table}[H]
\centering
\small
\tableformat
\setlength{\tabcolsep}{5pt}
\begin{tabular}{lccc}
\toprule
\rowcolor{vrBlueFill}
 & \textbf{QASPER} & \textbf{LitQA2} & \textbf{MOSAIC} \\
\rowcolor{vrBlueFill}
 & {\scriptsize(single-doc)} & {\scriptsize(multi-doc)} & {\scriptsize($\geq$5-paper)} \\
\midrule
\rowcolor{vrGrayFill}
\multicolumn{4}{l}{\textbf{\color{vrGray}Question composition}} \\
Benchmark questions & 300 & 54 & 49 \\
\;\; extractive / abstractive & 151 / 77 & --- & --- \\
\;\; yes-no / unanswerable & 43 / 29 & --- & --- \\
\;\; MCQ & --- & 54 & --- \\
\;\; multi-hop inference & --- & --- & 24 \\
\;\; aggregate / compare & --- & --- & 13 / 12 \\
\addlinespace[2pt]
\rowcolor{vrGrayFill}
\multicolumn{4}{l}{\textbf{\color{vrGray}Corpus coverage}} \\
Corpus papers & 416 & 61 & 124 \\
Gold papers (in study) & 215 & 54 & 124 \\
Publication years & --- & 2012--2024 & 2012--2026 \\
\addlinespace[2pt]
\rowcolor{vrGrayFill}
\multicolumn{4}{l}{\textbf{\color{vrGray}Document length}} \\
Paragraph units/paper (mean) & 49.6 & --- & --- \\
Paragraph units/paper (median; range) & 44.5; 7--309 & --- & --- \\
Physical pages/paper (mean) & --- & 21.8 & 12.8 \\
Physical pages/paper (median; range) & --- & 18; 1--57 & 12; 6--34 \\
\addlinespace[2pt]
\rowcolor{vrGrayFill}
\multicolumn{4}{l}{\textbf{\color{vrGray}Evidence}} \\
Gold evidence records/q (mean) & 2.5 & 1.0 & 5.1 \\
Gold papers/q (range) & 1 & 1 & 5--7 \\
\bottomrule
\end{tabular}
\caption{\textbf{Benchmark and corpus statistics.} Answer-type rows sum to the benchmark-question total. Evidence records differ in granularity: QASPER supplies gold paragraphs, LitQA2 supplies one key passage per question, and MOSAIC supplies multi-paper evidence with physical page locations.}
\label{tab:stats}
\end{table}

\FloatBarrier
\section{LitQA2 Evidence-Mapping Diagnostic}
\label{app:litqa-evidence}
LitQA2 supplies one gold document and one key passage per question rather than complete page-level annotations. Exact source-text matching mapped 10 of 54 key passages to physical pages. Table~\ref{tab:litqa-evidence} reports the resulting page-set scores for completeness, but these values are not evidence of corpus-level document precision and do not support the main component claims.

\begin{table}[H]
\centering
\small
\tableformat
\begin{tabular}{lcc}
\toprule
\rowcolor{vrBlueFill}
\textbf{Variant (agent)} & \textbf{Mapped-page recall} $\uparrow$ & \textbf{Mapped-page precision} $\uparrow$ \\
\midrule
\rowcolor{vrHighlight}
Full system & 0.151 & 0.076 \\
\addlinespace[2pt]
\rowcolor{vrGrayFill}
$-$ tree layer (search $\to$ pages) & 0.173 & 0.109 \\
$-$ vector layer (tree only, all docs) & 0.245 & 0.207 \\
\rowcolor{vrGrayFill}
$-$ relevance gates (raw RRF) & 0.151 & 0.108 \\
\bottomrule
\end{tabular}
\caption{\textbf{Exploratory LitQA2 page-mapping diagnostic.} Values are means over three seeds. Gold pages are available for only 10 of 54 questions after key-passage matching, and predictions are evaluated within the gold document. These values should not be compared directly with the QASPER or MOSAIC evidence metrics.}
\label{tab:litqa-evidence}
\end{table}

\FloatBarrier
\section{Query-Time Inference Cost}
\label{app:inference-cost}

Table~\ref{tab:inference-cost} reports the operational cost measurements separately from the main accuracy and evidence-localization results. VecTree-RAG used more query-time computation than Dense RAG, Dense RAG with reranking, and RAPTOR on QASPER and LitQA2. The comparison measures inference only; it does not include a corpus-matched accounting of embedding construction or serialized index size.

\begin{table}[H]
\centering
\small
\tableformat
\setlength{\tabcolsep}{4pt}
\begin{tabular}{lrrrr}
\toprule
\rowcolor{vrBlueFill}
 & \multicolumn{2}{c}{\textbf{QASPER}} & \textbf{LitQA2} & \textbf{MOSAIC} \\
\cmidrule(lr){2-3}\cmidrule(lr){4-4}\cmidrule(lr){5-5}
\rowcolor{vrBlueFill}
\textbf{System} & \textbf{Tokens/q} $\downarrow$ & \textbf{Wall (s)} $\downarrow$ & \textbf{Tokens/q} $\downarrow$ & \textbf{Tokens/q} $\downarrow$ \\
\midrule
Dense RAG & 7{,}660 & 11.0 & 9{,}696 & 194{,}187 \\
Dense RAG w/ Reranker & 7{,}722 & 20.2 & 8{,}660 & \phantom{0}16{,}648 \\
RAPTOR & \textbf{5{,}594} & 15.3 & 9{,}045 & \phantom{0}18{,}917 \\
Search-o1 & 31{,}261 & 25.5 & 141{,}173 & 409{,}748 \\
\addlinespace[2pt]
\rowcolor{vrHighlight}
VecTree-RAG (agent) & 41{,}186 & 45.9 & 127{,}872 & 279{,}337 \\
\bottomrule
\end{tabular}
\caption{\textbf{Inference cost across benchmarks.} Values are averaged over three seeds. Tokens per query are mean cumulative API-billed totals, and wall time is the QASPER mean. The MOSAIC Dense RAG value uses the expanded-context setting with $k=232$.}
\label{tab:inference-cost}
\end{table}

The LitQA2 component removals also changed query-time computation (Table~\ref{tab:ablation-cost}). Removing the vector layer produced the largest increase because the agent had to inspect trees across the full corpus.

\begin{table}[H]
\centering
\small
\tableformat
\begin{tabular}{lr}
\toprule
\rowcolor{vrBlueFill}
\textbf{LitQA2 variant (agent)} & \textbf{Tokens/q} $\downarrow$ \\
\midrule
\rowcolor{vrHighlight}
Full system & 127{,}872 \\
\addlinespace[2pt]
\rowcolor{vrGrayFill}
$-$ tree layer (search $\to$ pages) & 194{,}821 \\
$-$ vector layer (tree only, all docs) & 490{,}482 \\
\rowcolor{vrGrayFill}
$-$ relevance gates (raw RRF) & 104{,}215 \\
\bottomrule
\end{tabular}
\caption{\textbf{Query-time cost of the LitQA2 component ablations.} Values are mean cumulative API-billed tokens per query, averaged over three seeds.}
\label{tab:ablation-cost}
\end{table}

\FloatBarrier
\section{Indexing and Tree-Search Intrinsics}
\label{app:intrinsics}
The standalone verify--retry--fallback procedure is not part of the evaluated corpus-agent loop. We therefore report it as an intrinsic measurement. Across 61 indexed LitQA2 papers, the mean section-title verification rate increased from 0.910 to 0.924 after repair. The corresponding medians were 0.957 and 0.958, with 2.1 repair rounds on average. Papers averaged 21.8 pages, and the LLM indexing stages required 254\,s per paper on average, excluding Marker parsing.

On a 100-question QASPER sample, the rule-based node-confidence score was 1.0 for every query, and the lexical fallback was never triggered. This result shows that returned identifiers satisfied the validation rules. It does not measure whether the selected nodes contained the correct answer.

\section{Judge Robustness}
\label{app:judges}
DeepSeek-V4-Pro judged outputs generated by DeepSeek-V4-Flash, creating a possible model-family preference. Deterministic checks assess agreement only where answer extraction is mechanically unambiguous (Table~\ref{tab:judge-agreement}). These response-level diagnostic subsets were assembled from available evaluation runs and are not a separate re-evaluation of the five-system main comparison. They do not validate judge behavior on longer free-text responses.

\begin{table}[H]
\centering
\small
\tableformat
\begin{tabular}{lrrr}
\toprule
\rowcolor{vrBlueFill}
\textbf{Benchmark} & \textbf{Mechanical subset} & \textbf{Agreements} & \textbf{Agreement} \\
\midrule
QASPER yes/no & 387 & 383 & 0.990 \\
LitQA2 option selection & 470 & 469 & 0.998 \\
\bottomrule
\end{tabular}
\caption{\textbf{Agreement between the LLM judge and deterministic extraction on mechanically unambiguous outputs.} The response-level diagnostic subsets exclude outputs whose final answer cannot be extracted reliably by a fixed rule.}
\label{tab:judge-agreement}
\end{table}

The same judge rubric was applied to every system, which reduces one source of procedural asymmetry. However, systems differ in answer length and format, so symmetric scoring does not eliminate family or style bias. Evidence-page precision and recall are independent of the judge, but the headline answer scores are not. Regrading all outputs with multiple out-of-family judges remains the appropriate robustness test.

\FloatBarrier
\section{Prompt Templates}
\label{app:prompts}
This appendix transcribes selected query-time and grading prompts verbatim from the released code. Corpus-dependent fields (document counts, year ranges) and per-question fields are shown as \texttt{\{braced\}} placeholders filled at runtime. The complete prompt set is available in the released evaluation harness.

\paragraph{Agent system prompt.} The query-time ReAct agent is driven by the following prompt (\texttt{src/agent.py}, \texttt{build\_system\_prompt}). It is deliberately short ($\approx$400 tokens) and encodes the progressive-disclosure discipline.

\begin{lstlisting}[style=prompt]
You are a research assistant with access to a corpus of {total_documents} scientific papers ({year_range}) across {journal_count} journals. You answer questions by searching and reading these papers.

## Workflow
Classify the question, then act:
1. Library question (how many papers, what's available, list by author/year/topic):
   Answer from the corpus stats above when sufficient. Use search_documents("*") with
   filters to list or browse specific papers.
2. Research question (about content, methods, findings, comparisons):
   search_documents(query) -> pick relevant papers -> get_document_tree -> read summaries
   -> get_page_content for key sections -> synthesize answer with citations.
   If 0 results, reformulate with synonyms, broader terms, or author names.
3. Follow-up on papers already discussed:
   Use known doc_ids directly. Search again only if you need additional papers.

Some papers have supporting information (SI). Search results may include
related_documents with SI doc_ids -- use get_document_tree on those when needed.

## Token Budget Rules
- Read tree summaries BEFORE requesting pages.
- Never retrieve more than 5 pages without synthesizing first.
- If you have enough information to answer, stop retrieving.
- Use filters (year_min, year_max, journal, author) to narrow searches when appropriate.

## Citation Format
Cite as: (Author et al., Year, doc_id, p.X).

## Multi-Document Synthesis
When the question requires analysis across papers, synthesize with comparative language
and cite each source inline. If the question asks for a list or identification, a concise
answer is appropriate -- do not add unsolicited analysis sections.

## Images
Page text may contain inline image references like ![](filename.jpeg). Use get_page_image
to view figures and tables. Describe what the figure shows and incorporate it.
\end{lstlisting}

\paragraph{QASPER correctness judge.} Free-text answers are graded by the following judge (\texttt{eval/metrics/qasper\_regrade.py}), under the system message \emph{``You are a strict grader for free-text questions about scientific papers. Respond with JSON only.''}

\begin{lstlisting}[style=prompt]
Question: {question}

Reference answers (any one is acceptable):
{references}

A system produced the response below. Decide whether the response's FINAL answer
matches the meaning of ANY reference answer.
- "correct": conveys the same answer as a reference (for yes/no, the same polarity;
  if a reference is "Unanswerable", the response must conclude the question cannot be
  answered from the paper).
- "partial": overlaps a reference but misses or distorts a key part.
- "incorrect": contradicts every reference, or answers a different question.

Response:
"""{answer_text}"""

Reply with ONLY a JSON object: {"verdict": "correct|partial|incorrect",
"rationale": "<one short sentence>"}
\end{lstlisting}

\paragraph{LitQA2 MCQ judge.} Multiple-choice answers are graded by the following judge (\texttt{eval/\allowbreak metrics/\allowbreak mcq\_judge.py}), under the system message \emph{``You are a strict grader for multiple-choice questions about scientific papers. Respond with JSON only.''}

\begin{lstlisting}[style=prompt]
Question: {question}

Options:
{lettered_options}

The correct option is {gold_letter}. {gold_text}

A system produced the response below. Decide whether the response, as its FINAL answer,
selects the correct option. Merely mentioning or discussing the correct option's content
is NOT enough; it must be the response's conclusion. Use "abstain" only if the response
concludes there is insufficient information to answer.

Response:
"""{answer_text}"""

Reply with ONLY a JSON object: {"verdict": "correct|incorrect|abstain",
"selected": "<the option letter the response chose, or none>",
"rationale": "<one short sentence>"}
\end{lstlisting}

\paragraph{MOSAIC synthesis judges.} MOSAIC answer correctness follows the GraphRAG-Bench protocol (\texttt{eval/metrics/judge.py}): an LLM decomposes each answer into atomic statements, classifies each against the gold statements as true positive, false positive, or false negative to yield a factuality F1, and separately judges faithfulness by testing whether each answer statement is directly inferable from the read context. Answer correctness combines factuality F1 with scaled SPECTER embedding cosine as $0.75\,$F1$\,+\,0.25\,s_{\mathrm{cos}}$; coverage is the fraction of extracted reference facts attributed to the response. All MOSAIC sub-judgments use DeepSeek-V4-Pro at temperature~0.0, matching the QASPER and LitQA2 judge. The full prompt text for each sub-judge is released with the harness.

\bibliographystyle{plainnat}
\bibliography{references}

\end{document}